# An Energy-Efficient Resource Management System for a Mobile Ad Hoc Cloud

**Sayed Chhattan Shah, Senior Member, IEEE**

Department of Information Communication Engineering

Hankuk University of Foreign Studies, South Korea

Corresponding author: Sayed Chhattan Shah (e-mail: shah@hufs.ac.kr).

This research was supported by Hankuk University of Foreign Studies Research Fund of 2018 and Basic Science Research Program through the National Research Foundation of Korea funded by the Ministry of Science, ICT & Future Planning (2017R1C1B5017629).

**ABSTRACT** Recently, mobile ad hoc clouds have emerged as a promising technology for mobile cyber physical system applications, such as mobile intelligent video surveillance and smart homes. Resource management plays a key role in maximizing resource utilization and application performance in mobile ad hoc clouds. Unlike resource management in traditional distributed computing systems, such as clouds, resource management in a mobile ad hoc cloud poses numerous challenges owing to the node mobility, limited battery power, high latency, and the dynamic network environment. The real-time requirements associated with mobile cyber physical system applications make the problem even more challenging. Currently existing resource management systems for mobile ad hoc clouds are not designed to support mobile cyber physical system applications and energy-efficient communication between application tasks. In this paper, we propose a new energy-efficient resource management system for mobile ad hoc clouds. The proposed system consists of two layers: a network layer and a middleware layer. The network layer provides ad hoc network and communication services to the middleware layer and shares the collected information in order to allow efficient and robust resource management decisions. It uses (1) a transmission power control mechanism to improve energy efficiency and network capacity, (2) link lifetimes to reduce communication and energy consumption costs, and (3) link quality to estimate data transfer times. The middleware layer is responsible for the discovery, monitoring, migration, and allocation of resources. It receives application tasks from users and allocates tasks to nodes on the basis of network- and node-level information.

## I. INTRODUCTION

Recently, mobile ad hoc clouds have emerged as a promising technology for mobile cyber physical system (CPS) applications, such as mobile intelligent video surveillance and smart homes. In mobile CPS applications, various kinds of sensing and computing devices are deployed in order to monitor and analyze numerous situations and take the necessary actions in real time [1]. This involves sophisticated image and video processing algorithms, which require a vast amount of computing and storage resources. In order to address this issue, the conventional approach is to send the collected data to an application on an Internet cloud accessible through an infrastructure-based system, such as a cellular network [1]. This approach has several issues, such as high transmission energy consumption and communication latency. In addition, it cannot be used if preexisting network infrastructure is not available. In order to overcome the drawbacks of the conventional cloud-based approach, a new technology called mobile ad hoc clouds is currently being developed, in which multiple mobile devices, interconnected through a mobile ad hoc network, are combined to create a virtual supercomputing node or a small local cloud data center [1]. Mobile ad hoc clouds not only are deployable in network environments with no infrastructure but also significantly reduce the transmission energy consumption and communication latency.

A block diagram and the architecture of a mobile ad hoc cloud computing system are shown in Figs. 1 and 2. Nodes in mobile ad hoc clouds are divided into three categories: service master nodes (SMNs), service provider nodes (SPNs), and service consumer nodes (SCNs). SMNs are responsible for allocating tasks submitted by the SCNs to the SPNs on the basis of the resource allocation policy. Nodes communicate with each other through a mobile ad hoc network, which provides several communication services, including routing and medium access. The mobile cloud middleware layer is responsible for resource management, failure management, mobility management, communication management, and task management. In addition, the middleware hides all the complexities and provides a single system image to the user and applications running on the system.

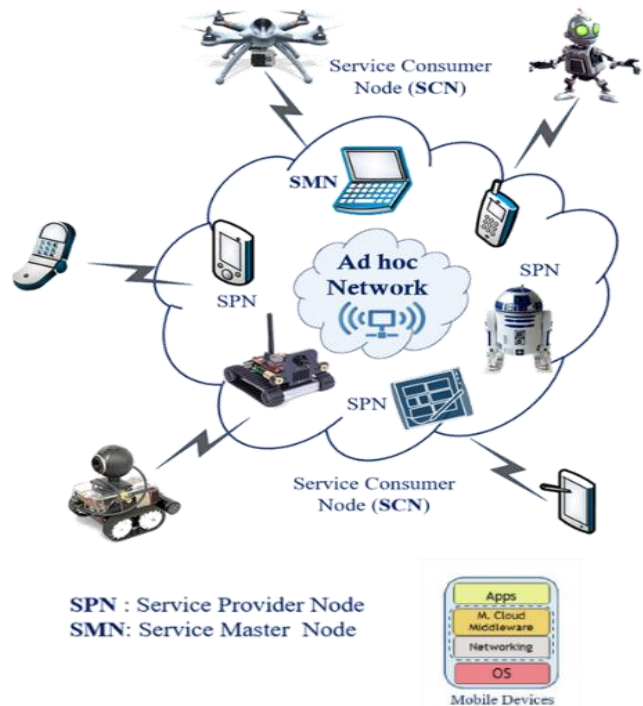

**Figure 1.** Block diagram of a mobile ad hoc cloud computing system.

A resource management system (RMS) is an integral part of any distributed computing system and is responsible for the discovery, monitoring, and allocation of network and system resources. Compared to traditional parallel and distributed computing systems, such as clouds, resource management in a mobile ad hoc cloud poses numerous challenges owing to the node mobility, limited battery power, high latency, and the dynamic network environment [1]. The real-time requirements associated with mobile CPS applications make the problem even more challenging. Research into resource management for mobile ad hoc clouds is still in a preliminary phase, and very few schemes based on a decentralized architecture have been proposed to address issues such as node mobility, energy management, and task failure [1].

In this paper, we propose an energy-efficient RMS for a mobile ad hoc cloud. The proposed system consists of two layers: a network layer and a middleware layer. The network layer provides ad hoc network and communication services to the middleware layer and shares the collected information in order to allow efficient and robust resource management decisions. It uses (1) a transmission power control mechanism to improve energy efficiency and network capacity, (2) link lifetimes to reduce communication and energy consumption costs, and (3) link quality to estimate data transfer times. The middleware layer is responsible for the discovery, monitoring, migration, and allocation of resources. It receives application tasks from users and allocates tasks to nodes on the basis of network- and node-level information.

Compared to already existing RMSs, the proposed system focuses on mobile CPS applications and energy-efficient communication between tasks. In addition, it aims to address task failure by adopting a failure avoidance mechanism. In order to make effective resource allocation decisions, a cross-layer design approach is used. Our research on transmission power control mechanisms is used to reduce transmission power energy consumption and increase concurrent transmissions in the network. The new system considers link quality and lifetime to reduce data transfer costs and communication energy consumption.

The key contributions of this paper are as follows:
- An energy-efficient RMS for mobile ad hoc clouds
- A network layer to provide energy-efficient and reliable ad hoc network and communication services

- A link lifetime aware routing protocol based on a transmission power control mechanism to reduce the transmission energy consumption and data transfer costs
- An efficient distance-based discovery mechanism to reduce the communication costs
- A Markov-chain-based LLPH model to predict link lifetimes
- A data transfer time estimation model that considers packet overhead, the number of dropped and lost packets, and traffic at neighboring nodes
- An energy consumption estimation model
- A task completion time estimation model

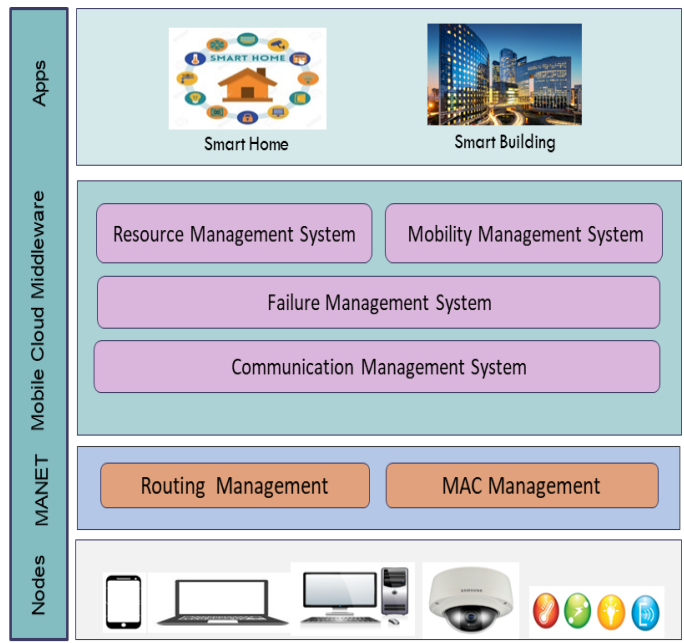

**Figure 2.** Architecture of a mobile ad hoc cloud computing system.

The rest of the paper is organized as follows. Section II focuses on the related work, Section III describes energy-efficient mobile ad hoc cloud RMS, Section IV describes the preliminary results, and Section V presents the conclusion.

## II. LITERATURE REVIEW
Research into resource management for mobile ad hoc clouds is still in a preliminary phase, and very few schemes based on a decentralized architecture have been proposed in order to address issues such as node mobility, energy management, and task failure. This section is divided into two parts. The first part describes the work related to resource allocation, and the second part focuses on resource discovery and monitoring schemes.

### A. RESOURCE ALLOCATION SCHEMES
A resource allocation scheme based on a distributed architecture was proposed in [2]. The scheme uses proactive and reactive failure management approaches and supports redundant execution of tasks to handle a failure. The scheme that was developed in [3] uses a program-controlled migration technique to address task failures. In order to allocate tasks, it uses a manager–worker model. Energy minimization and grid utility optimization problems are addressed in [4], and a scheme to reduce energy consumption was described in [5]. In order to balance the computational workload and energy consumption, the authors of [6] proposed a scheme that investigates the cooperation among mobile nodes. The scheme proposed in [7] addresses load balancing and scalability problems. In order to allocate a task, it uses a delayed response mechanism in which powerful nodes reply earlier than less powerful ones. A node mobility problem is addressed in [8] by profiling the mobility patterns of users. The authors of [9] developed several online and batch scheduling schemes to offload CPU-intensive tasks onto a mobile ad hoc cloud. The MinHop scheme assigns tasks to a node on the basis of the number of hops, whereas the minimum execution time (MET) with communication scheme assigns a task to a node that would take the minimum time to execute. The minimum completion time (MCT) with communication scheme selects a node with the minimum expected completion time. MinMinComm and MaxMinComm, which are based on traditional MET, MCT, MinMin, and MaxMin schemes, have been proposed for batch scheduling. Compared to traditional schemes, in the proposed scheme, communication costs are considered. Task allocation problem in numerous network environments have been studied in [10]. A separate scheme has been proposed for each environment. A greedy scheme has been developed for an ideal network environment, which assigns a task to the node with the minimum task completion time.

Numerous job-stealing schemes (e.g., random, best rank aware, and worst rank aware) were proposed in [11]. These schemes are based on a centralized architecture and are aimed at reducing the energy consumption. The problems of a dynamic and unpredictable network environment, energy consumption, and node failure are addressed in [12]. The scheme proposed in [13] addresses the uncertainty problem using application waypoints. A node executing a task sends an estimate of the residual task completion time to a broker node. If the broker node fails to receive a task progress update at a specified waypoint, it assumes that the node has failed and takes necessary measures to complete the task.

In order to address the transmission energy consumption problem, an energy-efficient resource allocation scheme was proposed in [14]. This scheme uses a transmission power control mechanism to reduce transmission energy and a hybrid architecture for making efficient decisions. The node mobility problem was addressed in [15]. The scheme proposed in [15] uses the history of user mobility patterns, task characteristics, and distance information to reduce data transfer times. An efficient and robust resource allocation scheme that was developed in [16] aims to reduce task completion times and transmission energy consumption using a transmission power control mechanism and user mobility patterns.

Existing resource allocation schemes for mobile ad hoc systems do not consider the network environment, task queue size, or CPU overhead. Very few schemes consider the network environment, but they do not consider link quality, link lifetime, or the migration or reallocation of tasks.

This work is different from the schemes proposed in the literature, because it focuses on mobile CPS applications and energy-efficient communication between tasks. In addition, it aims to address task failure by adopting a failure avoidance mechanism. In order to make effective resource allocation decisions, a cross-layer design is used. Our previous research into transmission power control mechanisms is used to reduce the transmission energy consumption and increase concurrent transmissions in the network. The new system makes allocation decisions on the basis of CPU speed, task queue size, CPU overhead, link quality, and link lifetime.

Table I. Summary of resource management schemes.

| | Transmission Energy Aware | Processing Energy Aware | Mobile CPS App Support | Cross Layer Design | Network Aware | Mobility Management | Load Balancing | Failure Management | Link Lifetime Aware | Link Quality Aware |
|---|---|---|---|---|---|---|---|---|---|---|
| **Existing Schemes** | [14] [16] | [4] [5] [6] [11] [13] | None | None | [9] [14] [15][16] | [8] [15] [16] | [6] [7] [11] | [2] [3] [12] [13] [15] [16] | None | None |
| **Proposed Scheme** | √ | | √ | √ | √ | √ | | √ | √ | √ |

## B. RESOURCE DISCOVERY AND MONITORING

Numerous resource discovery protocols have been developed for wired and wireless networks and systems. In [17, 28, 30], a survey of discovery technologies and a categorization based on search strategies and protocols are provided. The first category includes technologies such as Bluetooth Low Energy, which enables the discovery of resources in close spatial proximity. The client node sends a discovery message or may wait for an advertisement message. The second category includes technologies and protocols that enable the discovery of a resource on the network [19–21]. In order to discover a resource, a client multicasts a discovery request message over the network. A host receives the discovery message and replies via a response message. The third category uses a central or distributed directory [22–24], which stores information about resources. The information is registered by resource providers. A discovery request message is sent to the directory, which then replies via a discovery response message. In [18], a resource discovery and selection process based on matching parameters was proposed. The authors used an analysis matrix to describe multiple criteria in a decision analysis problem. The behavior of simple additive weighting and the TOPSIS and VIKOR multiobjective decision methods is analyzed. Normalization, the Euclidean distance, and sorting algorithms are used to optimize the resource selection process. The authors of [25] proposed a resource discovery mechanism in order to enhance the search efficiency over the social Internet of Things. The mechanism is based on preference and movement pattern similarity. A three-layer network structure is adopted, in which each node has some common interest. Based on these interests, subcommunities and global communities are formed. Every node has a preference vector, which is used to calculate the similarity between two nodes. Different parameters, such as weighting parameters, user-defined parameters, location coordinates, temporal features, and overlapping movement regions, are utilized. In order to address the challenges of the distributed nature, energy efficiency, scalability, and fast discovery, the authors of [26] proposed a device-to-device discovery protocol. This protocol allows neighboring nodes to form a group and to be synchronized. The nodes use table and time interval information in the group for discovery and beacon broadcasting. A device discovery approach and a connection establishment scheme for opportunistic networks were developed in [27], where a device announces its existence using a beacon stuffing method. The authors also proposed a score-based scanning schedule for device discovery to reduce the energy consumption.

The resource discovery protocols that were developed for wired and infrastructure-based wireless networks cannot be used in mobile ad hoc systems, which are characterized by a dynamic and unreliable network environment. Wired networks have stationary devices, whereas infrastructure-based wireless networks have nodes with restricted mobility. The resource discovery protocols that were developed for mobile ad hoc systems work at either the application layer or the network layer. The application layer protocols focus on the architecture and the distribution of discovery components and are not aware of the communication mechanisms employed at the network layer [28]. They also do not consider node mobility and the dynamic network environment. Network layer discovery protocols are developed as part of a routing protocol and are used to discover limited network-level information, such as the node ID and number of hops [29].

## III. AN ENERGY EFFICIENT RESOURCE MANAGEMENT SYSTEM FOR A MOBILE AD HOC CLOUD

The architecture of an energy-efficient mobile cloud RMS is shown in Fig. 3. The RMS is divided into two layers: a network layer and a middleware layer. The network layer provides network and communication services and also collects network-level information, such as link quality and lifetime, which is used by the middleware layer during the resource allocation process. The key components of each layer are described below.

### A. RMS NETWORK LAYER

The network layer provides ad hoc network [45] and communication services to the middleware layer, and it shares the collected information in order to enable efficient and robust resource management decisions.

The network layer also plays a key role in the overall network and system performance. The key factors that affect the performance of the network layer are the transmission power, link lifetime, and link quality. The transmission power affects the transmission energy consumption and network capacity. The link quality determines the data transfer time, whereas the link lifetime determines the communication and energy consumption costs. Numerous protocols have been proposed in the literature, but they consider either the transmission power [38–43], link lifetime [44–49], or link quality [36, 37]. Transmission power aware routing protocols have poor discovery mechanisms. There are two categories of protocols based on the link lifetime. The first category includes protocols that rely on the current information about nodes, such as the location, distance, speed, residual energy, and transmission energy consumption. The schemes in the second category exploit historical information about nodes and users, such as mobility patterns and social relationships. Link quality aware routing protocols do not consider the packet overhead, the number of dropped and lost packets, or the traffic at neighboring nodes.

In this paper, an energy-efficient and link lifetime aware network layer is proposed in order to improve the energy efficiency and data transfer times. The proposed layer uses (1) a transmission power control mechanism to improve the energy efficiency and network capacity, (2) link lifetimes to reduce the communication and energy consumption costs, and (3) link quality to estimate data transfer times. The proposed process at the network layer is as follows:

- Receive a data transmission request from the middleware.
- Estimate the data transfer time.
- Predict the link lifetime.
- Select an energy-efficient route with a link lifetime greater than or equal to the estimated data transfer time.

The architectural elements of the network layer are described below.

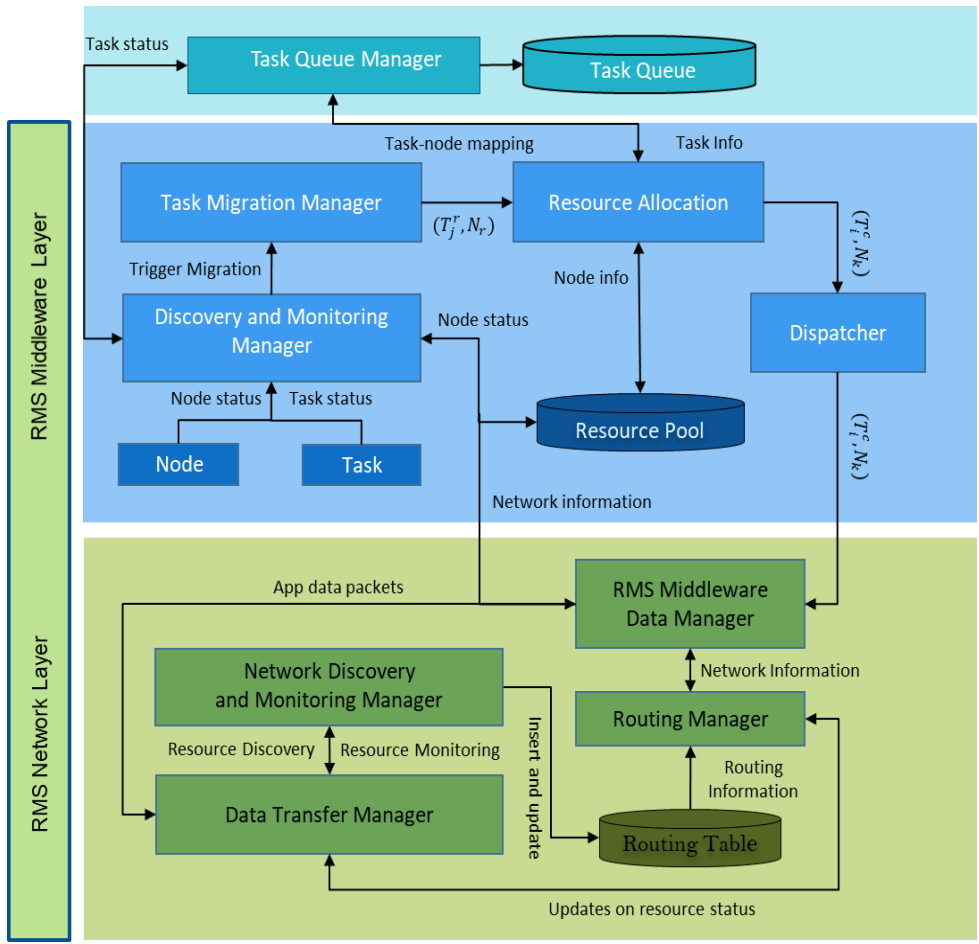

**Figure 3:** RMS architecture.

#### 1) MIDDLEWARE DATA MANAGER
The middleware data manager hides the network's complexity and provides an interface to the middleware for data transmission across the network.

#### 2) NETWORK DISCOVERY AND MONITORING MANAGER
The network discovery and monitoring manager discovers and monitors the following network-level information:
1) Node ID
2) Average number of dropped and lost packets
3) Available bandwidth
4) Neighboring traffic
5) Self-traffic
6) Total bandwidth
7) Energy consumption per packet
8) Received signal strength intensity

The information collected by the discovery and monitoring manager is used to estimate the data transfer times and predict link lifetimes. The following sections describe the discovery and communication mechanism and the process of estimating the data transfer times and link lifetimes.

*a) An Energy-Efficient Discovery and Communication*
In order to reduce the transmission energy consumption, a ClusterPow discovery and routing protocol was developed in [32]. This routing protocol significantly reduces the energy consumption but introduces a communication overhead owing to its poor discovery mechanism [16]. In this section, an efficient discovery mechanism is proposed to reduce this communication overhead. It is assumed that each node can transmit at multiple transmission power levels.

For each transmission power level $TPL_i$ (Fig. 4), each node maintains a separate routing table $RTP_i$. In order to discover nodes at each TPL, a source node uses the maximum transmission power to broadcast a discovery request packet. The node that receives the discovery request packet calculates the distance to the source node using the following equation:

$$D_{ij} = 10 \wedge \frac{(TX_{power} - RSSI)}{10 \times n}, \qquad (1)$$

where $D_{ij}$ is the distance between nodes $N_i$ and $N_j$, $TX_{power}$ is the transmission power of $TPL_i$, RSSI is the received signal strength intensity, and $n$ is a constant that depends on environmental factors.

Based on the distance, the node determines the transmission power level required to communicate with the source node. The node that received the discovery request packet then sends a reply packet at the selected transmission power level, which is also included in the packet.

When the source node receives the reply packet, it adds a new entry or updates an existing entry in the routing table. A routing entry comprises the node address and the transmission power level included in the reply packet.

In order to forward a packet to a destination, the routing manager accesses the lowest power routing table $RTP_{min}$ in which the destination is present and forwards the packet at power level $TPL_{min}$ to the next hop indicated in the routing table $RTP_{min}$.

To reduce communication costs, the discovery process should be started at a master node rather than at every node in the network.

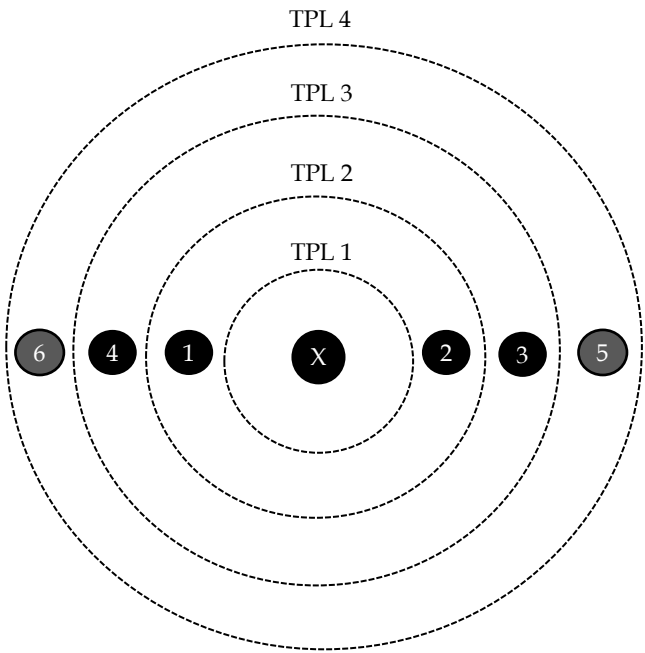

**Figure 4. Illustration of the transmission power control mechanism.**

*b) Estimating the Data Transfer Time*

The link quality reflects the available bandwidth and is used to estimate the data transfer times. In order to estimate the link quality, Model 2 that was developed in [36, 37] is used. Compared to existing work, the model in [36, 37] considers self-traffic and traffic at neighboring nodes:

$$LQ_{mk} = B_{available}(N_{mk}) = B_{channel} - \sum_{k \in N_{neighbors}(N_m)} B_{self}(N_k), \quad (2)$$

where $LQ_{mk}$ is the link quality between nodes $N_m$ and $N_k$, $B_{self}(N_k)$ is the total traffic between node $N_k$ and its neighbors, $B_{channel}$ is the total bandwidth, and $B_{available}(N_{mk})$ is the available bandwidth between nodes $N_m$ and $N_k$.

The application or middleware layer sends a data transmission request, which includes the data size, to the network layer. Based on the data size, the network layer estimates the data transfer time using Model 3,

$$E_{DTT} = (P_{kts} \times Pkt_{size}) \div LQ_{mk} + (P_{kts\_AvgDL} \times Pkt_{size}) \div LQ_{mk} \quad (3)$$

$$\text{Number of packets } P_{kts} = Data_{size} \div (Pkt_{size} - Pkt_{header\_size}) \quad (4)$$

where $P_{kts\_AvgDL}$ is the average number of dropped and lost packets, $Data_{size}$ is the quantity of data to be transmitted, $Pkt_{size}$ is the packet size, and $Pkt_{header\_size}$ is the packet header size.

Compared to existing work, the proposed model considers the packet overhead, the average number of dropped and lost packets, and the traffic at neighboring nodes. After estimating the data transfer time, the routing manager selects a route with an estimated lifetime greater than or equal to the estimated data transfer time. If multiple routes are available, then the route with the maximum probability is selected. For real-time data transmission, multiple routes can be selected to transfer data simultaneously to a destination node.

*c) Estimating the Link Lifetime*

In a previous study, we proposed a Markov-chain-based node location prediction scheme [31], which exploits the history of a user's mobility patterns to predict the next location. In this paper, a Markov-chain-based link lifetime estimation scheme is proposed. The main idea is to predict the available link lifetime on the basis of the history of link lifetime intervals. A route with an estimated lifetime greater than or equal to the application data transfer time is then selected for data transmission.

In order to predict the link lifetime, a Markov chain model is used. A Markov chain is a sequence of random variables $X_1, X_2, X_3, ...$ with the Markov property that, given the present state of the system, the future is independent of its past.

Formally,
$$\Pr(X_{n+1} = x | X_1 = x_1, X_2 = x_2, ..., X_n = x_n) = \Pr(X_{n+1} = x | X_n = x_n)$$

The random variable $X_i$ takes values from a countable discrete set S. The elements of S are called states, and S is referred to as the state space.

The underlying Markov model represents each state as a link lifetime interval, and transitions represent the possible link lifetime intervals that follow the current interval. Whenever a link becomes available, the Markov transition probability matrix, which is used to describe the transitions, is updated. A row in the transition matrix includes the probabilities of link lifetime intervals. In order to make a prediction, the algorithm scans the row of the transition matrix corresponding to the current state or link lifetime interval and selects the entry with the maximum probability. The maximum probability interval is the interval that has most frequently followed the current interval in the past.

As a demonstration, Fig. 5 shows three link lifetime intervals: short, medium, and long. A transition probability matrix is also shown in Fig. 5. For each interval, the transition probability matrix contains the probabilities of moving from that interval to another. For example, a link $L_{mn}$ with current state S or a short lifetime interval has a 0.5 probability that it will be available for S or a short duration, a 0.4 probability that it will be available for M or a medium duration, and a 0.1 probability that it will be available for L or a long duration.

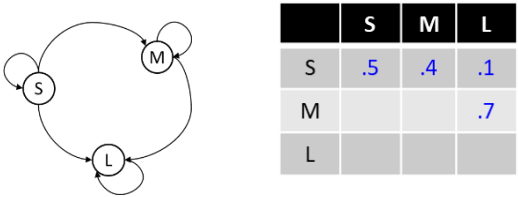

**Figure 5.** State transition diagram and transition probabilities.

A transition probability matrix is maintained for each link. The prediction is made once the link becomes available. The possible transitions are $S \to S, S \to M, M \to M$, and $M \to L$. The following transitions are not logical and, therefore, are not permitted: $M \to S, L \to M$, and $L \to S$.

The main objective of predicting the link lifetime and considering it during link, route, and node selection is to reduce the energy consumption, data transfer time, and thus CPS application completion time. The proposed model can easily be adopted in vehicular ad hoc networks.

### 3) DATA TRANSFER MANAGER

The data transfer manager is responsible for data transmission using an underlying communication technology such as Wi-Fi Direct. The proposed network layer is independent of the communication technology. It can be modified to support one or several communication technologies, such as Wi-Fi or ZigBee.

### 4) ROUTING TABLE

A routing table maintains multiple routes to a destination. A routing entry in a routing table includes the following information:
1) Next node
2) Destination node
3) Average number of dropped and lost packets
4) Link quality
5) Link lifetime

### 5) ROUTING MANAGER

In order to reduce the transmission energy consumption, a transmission power control mechanism is used. It is assumed that every node can transmit at multiple transmission power levels. For each transmission power level $TPL_i$, each node maintains a separate routing table $RTP_i$. In order to discover nodes, the discovery mechanism described in the previous section is used. Based on the discovered information, the link quality and link lifetime for each node accessible at each transmission power level are determined. The procedure of estimating the link quality and lifetime was described in the previous section.

In order to forward data to a destination, a source node accesses the lowest power routing table $RTP_{min}$ and selects a route whose lifetime is greater than or equal to the estimated data transfer time. If there are possible multiple routes, the route with the maximum probability is selected. The route selected from $RTP_{min}$ can improve the energy efficiency and network capacity. This is because communication at lower transmission power reduces the transmission energy consumption and increases the number of parallel transmissions and therefore the network's capacity. If no route exists to the destination in the lowest power routing table $RTP_{min}$, then a routing table for the next transmission power level $RTP_{min+1}$ is accessed. The route selection mechanism is described below.

An alternative approach is to calculate the weight of each route on the basis of the link quality, link lifetime, and energy consumption and then to select the route with the maximum weight.

### B. RMS MIDDLEWARE LAYER

Middleware layer is responsible for the discovery, monitoring, migration, and allocation of resources. It receives application tasks from users and allocates tasks to nodes on the basis of network- and node-level information. Each element of the middleware layer is described below.

---

Receive data transmission request

Previously estimated data transfer time $E_{DTT\_previous}$ = INFINITY

```
Probability of route lifetime P_Route_LT_Previous = 0
RTP_{i=1}
For each routing table RTP_i
  For each route in routing table RTP_i
    Estimate data transfer time E_DTT using data transfer time estimation model
    If ( ∃ a route with link lifetime Route_LT ≥ Estimated data transfer time E_DTT) and
    (E_DTT < E_DTT_previous) and (probability of route lifetime P_Route_LLT > P_Route_LT_Previous)
      Select a route for data transmission
      E_DTT_previous = E_DTT
      P_Route_LT_Previous = P_Route_LLT
      RTP_{i++}
```

**Algorithm 1:** A route selection mechanism.

### 1) TASK QUEUE MANAGER
The task queue manager manages the task queue. It receives tasks from users and inserts them into the task queue, and it also collects task results upon successful completion and communicates them to the users.

### 2) TASK QUEUE
The task queue has information about tasks and allocated nodes:

1) Task ID
2) Task type
3) Code size
4) Input data size
5) Output data size
6) Node ID
7) Task status
8) Task progress

### 3) RESOURCE POOL
The resource pool maintains the characteristics of each member of the mobile ad hoc cloud. It also maintains network-level information, such as the link quality and lifetime discovered and monitored by the network layer:

1) Destination node ID
2) Clock cycles per instruction
3) Clock cycle time
4) CPU energy consumption
5) System overhead
6) Task queue waiting time
7) Available memory
8) Available battery power
9) Transmission power level
10) Link quality
11) Link lifetime
12) Average number of dropped and lost packets

### 4) DISCOVERY AND MONITORING MANAGER
The discovery and monitoring manager integrated with the network layer is divided in two subcomponents, namely, node discovery and monitoring and network discovery and monitoring. The network discovery and monitoring manager is part of the network layer and is discussed in a related section.

The node discovery and monitoring manager discovers and monitors mobile devices in the communication range and tasks being executed on mobile nodes. The information discovered about nodes is stored in a resource pool, and tasks are stored in the task queue. Both the resource pool and the task queue are regularly updated by the monitoring manager, which triggers the migration manager under predefined conditions, such as node overutilization or underutilization or poor progress of a task.

The node discovery and monitoring manager has a distributed architecture. It is deployed on each node of the mobile ad hoc cloud. In order to discover and monitor resources, the discovery and monitoring manager on each node communicates the following messages with neighboring nodes.

#### a) Node Information Message
A node information message (NIM) is used to discover and monitor nodes in the coverage area. It is also used to keep track of nodes in or out of the coverage range. Each node broadcasts a NIM every x intervals. The node that receives a NIM adds a new entry or updates an existing entry in a resource pool. If a node does not receive a NIM from a member node during m intervals, the member node is declared as unavailable in the resource pool.

| Message type | Node ID | CPI | CCT | Broadcast ID |
|---|---|---|---|---|

Node information message

#### b) Node Information Update Message
A node information update message (NIUM) is used to communicate dynamic information about a node, such as task queue size and available memory. In order to reduce the communication overhead, each node broadcasts a NIUM when the task queue size, available memory, or available battery power crosses a threshold. To communicate dynamic information to new nodes, the following mechanism is used.

On receiving a NIM:
If (node ID in the NIM does not exist in the resource pool)
  Add a new entry to the resource pool
  Send a node dynamic information request message (NIRM)
  A node receiving a NIRM sends a NIUM

| Message type | Source ID | Destination ID |
|---|---|---|

Node information request message

| Message type | Node ID | Queue waiting time | Available memory | Battery power | Destination node ID |
|---|---|---|---|---|---|

Node information update message

### c) Members Information Message

A members information message (MIM) is used to share resource pool information with neighboring nodes. Each node periodically broadcasts a MIM.

| Message type | Node ID | Member ID | CPI | CCT |
|---|---|---|---|---|
| Member ID | CPI | CCT | Broadcast ID | |

Members information message

In order to share dynamic information about member nodes, each node broadcasts a member information update message (MIUM) when the task queue size, available memory, or available battery power crosses a threshold. To communicate dynamic information to new nodes, the following mechanism is used.

On receiving a MIM:
If (member ID in MIM does not exist in the resource pool)
  Add a new entry to the resource pool
  Send members dynamic information request message
  A node receiving an MDIRM sends a MIUM

| Message type | Source ID | Member ID |
|---|---|---|
| Member ID | Member ID | Destination ID |

Members dynamic information request message

| Message type | Source ID | Member ID | Queue waiting time |
|---|---|---|---|
| Available memory | Available battery power | Destination or broadcast ID | |

Members information update message

### d) Task Information Message

A task information message (TIM) is used to communicate the list of allocated or executing tasks at a node and the status of each task. This information is used by the resource allocation system to migrate or reallocate a task to another node in order to improve resource utilization, task performance, or energy efficiency. A TIM can also be used to detect task failure. If a node does not receive a TIM during $z$ intervals, it will assume the failure of a task and can take the necessary action. A TIM is sent every $z$ intervals to every node that has been sent the task for execution. If the task list cannot fit in a single message, then multiple messages are sent. With a centralized architecture, TIMs are sent to the node hosting the resource allocation service.

| Message type | Node ID | Task ID | Task status |
|---|---|---|---|
| Task ID | Task status | Destination ID | |

Task information message

### 5) RESOURCE ALLOCATION SERVICE

The resource allocation service assigns independent application tasks to nodes in mobile ad hoc clouds (Algorithm 2). It accesses task information from the task queue and node information from the resource pool and finds a suitable node that fulfills the task requirements. During matching, it considers the CPU speed, queue size, link quality, link lifetime, and transmission power required to send the data. The selected node ID and task ID are sent to the dispatcher, which then dispatches the selected task to the selected node.

Here is the step-by-step process for allocating a task (Fig. 6):

(1) A SCN submits task information to the task queue manager deployed on the SMN.

(2) The resource allocation service deployed on the SMN selects a node to execute the task.

(3) The ID of the selected node is sent to a dispatcher deployed on the SCN.

(4) The task code and related data are sent to a selected SPN.

(5) Upon successful completion of the task, the result is sent to the SCN.

(6) The task status is updated in the task queue deployed on the SMN.

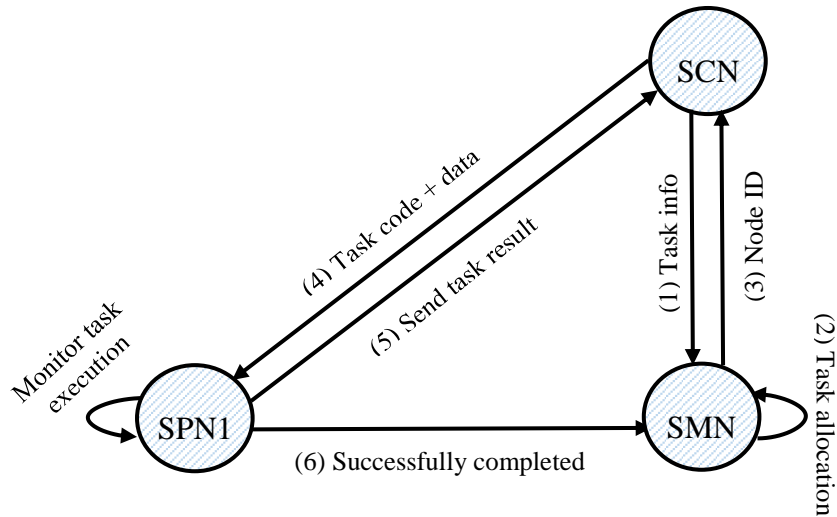

Figure 6. **Step-by-step process for allocating a task.**

Cost estimation models and the resource allocation algorithm that selects a node to execute a task are described below.

*Cost Estimation Models*

*Estimated completion time ($E_{CT}$) of task $T_i$ on the node $N_k$*

The key factors that contribute to the task completion time are the task execution time ($E_{ET}$) and the data transfer time ($E_{DTT}$). In order to estimate the completion time of task $T_i$ on node $N_k$, we use

$$E_{CT}(T_i, N_k) = E_{ET}(T_i, N_k) + E_{DTT}(T_i, N_k). \quad (5)$$

The process of estimating the execution time ($E_{ET}$) of task $T_i$ on node $N_k$, is as follows.

The task execution time consists of the CPU processing time and task queue time, which is defined as the time that a task waits in a queue. In order to estimate the execution time ($E_{ET}$) of task $T_i$ on node $N_k$, we use

$$E_{ET}(T_i, N_k) = E_{PT}(T_i, N_k) + E_{QT}(T_i, N_k), \quad (6)$$

which considers the CPU processing time ($E_{PT}$) and the task queue time ($E_{QT}$).

The CPU processing time ($E_{PT}$) of task $T_i$ on node $N_k$, is estimated using

$$E_{PT}(T_i, N_k) = I_i \times N_{CPI}^k \times N_{CCT}^k, \quad (7)$$

where $I_i$ is the number of instructions in task $T_i$, $N_{CPI}^k$ are the clock cycles per instruction of node $N_k$, and $N_{CCT}^k$ is the clock cycle time of node $N_k$.

In order to estimate the task waiting time at queue, we use

$$E_{QT}(T_i, N_k) = E_{PTE}(T_i, N_k) + E_{PT\_TQ} + ((m+2) \times \phi), \quad (8)$$

$$E_{PTE}(T_i, N_k) = (I_i - IE_i) \times N_{CPI}^k \times N_{CCT}^k, \quad (9)$$

where $\phi$ is the time spent by the system in allocating and deallocating tasks to a CPU and $IE_i$ is the number of processed instructions.

The CPU processing time for tasks waiting in a queue is estimated through

$$E_{PT\_TQ} = \sum_{n=1}^{m} E_{PT}(T_n, N_k), \quad (10)$$

which also considers the estimated execution time of tasks waiting for execution on the CPU. Here $m$ is the number of tasks in the queue.

*Estimated Data Transfer Time ($E_{DTT}$)*
The data transfer time is estimated using

$$E_{DTT} = (P_{kts} \times Pkt_{size}) \div LQ_{mk} + (P_{kts\_AvgDL} \times Pkt_{size}) \div LQ_{mk} \quad (11)$$

where $P_{kts}$ is the number of packets, $Pkt_{size}$ is the packet size, $LQ_{mk}$ is the link quality between nodes $N_m$ and $N_k$, and $P_{kts\_AvgDL}$ is the average number of dropped and lost packets. For details, refer to Section III.A.2.

*Estimated Energy Consumption ($E_{EC}$)*
In order to estimate the energy consumption, we use

$$E_{EC}(T_i, N_k) = (\alpha \times E_{PT}(T_i, N_k)) + (\beta \times P(T_i)), \quad (12)$$

where α is the CPU energy consumption per unit of time and β is the wireless channel energy consumption per transmitted packet.

The equation considers CPU energy consumption and communication energy consumption. CPU energy consumption is estimated using [50]

$$\alpha = P_{static} + P_{dynamic}, \quad (13)$$

where the dynamic CPU power consumption is

$$P_{dynamic} = A \times C \times V^2 \times F, \quad (14)$$

where A is the number of active logic gates, C is the total capacitance load, V represents the voltage, and F represents the frequency.

---

**Input:** Set of tasks $T_i \in T$ in the task queue
  Set of nodes $N_k \in N$ in the resource pool
  Network information stored in $RTP_i$ at the network layer
**Output:** Assignment of task $T_i$ to service provider node $N_{spn} \in N$

1  Repeat until task queue is not empty
2  {
3  For each task $T_i \in T$ submitted by SCN node $N_C$
4   {
5    Previously estimated task completion time $E_{CT}(T_i, N_{pre}) = 0$
6    $RTP_{i=1}$
7    Probability of route lifetime $P\_Route_{LLT\_previous} = 0$
8    Service provider node $N_{spn}$ = null
9    For each node $N_k \in N$ in routing table $RTP_i$
10   {
11    For each route to $N_k$ in routing table $RTP_i$
12    Estimate task completion time using task completion time estimation model:
      $E_{CT}(T_i, N_k) = E_{ET}(T_i, N_k) + E_{DTT}(T_i, N_k)$
13    Estimate processing energy consumption:
      $E_{EC}(T_i, N_k) = (\alpha \times E_{PT}(T_i, N_k))$
14    If ( ∃ a route to $N_k$ with lifetime $Route_{LT} \geq$ Estimated data transfer time $E_{DTT}(T_i, N_k)$ &
       ($E_{CT}(T_i, N_k) < E_{CT}(T_i, N_{pre})$) & Probability of route lifetime $P\_Route_{LLT} > P\_Route_{LLT\_previous}$
15     Assign $T_i$ to $N_k$
16     $E_{CT}(T_i, N_{pre}) = E_{CT}(T_i, N_k)$
17     $P\_Route_{LLT\_Previous} = P\_Route_{LLT}$
18    }
19    $RTP_{i++}$
20   }
21   $N_{spn} = N_k$
22   Send task-node assignment $(T_i, N_{spn})$ to service consumer node $N_C$
23   Update the task queue and resource pool with task-node assignment $(T_i, N_{spn})$ and
     completion time $E_{CT}(T_i, N_{spn})$
24  }

**Algorithm 2:** The proposed resource allocation algorithm.

---

6) DISPATCHER

A dispatcher is deployed on each node. It performs several functions, including the following:

(1) It dispatches task code and data to a node selected by the resource allocation service.
(2) It interfaces with the local execution and operating environment to enable the execution of a received task.
(3) Upon successful completion of the task, it sends the result to the SCN and status in the resource pool and the task queue.

#### 7) TASK MIGRATION

A task migration process is initiated for the following events:

(1) A node is about to fail owing to mobility or low battery power.
(2) A node is underutilized or overutilized.
(3) A more powerful or suitable node joins the network.

The migration process consists of three phases (Fig. 7).

(1) *Task Migration Setup Phase*. The task migration service must decide which task to migrate and where to migrate it to. In order to find a suitable target node, the task migration service uses the resource allocation service. Simultaneously, a task selected for migration is cloned at a source node. The resource allocation service sends the ID of the selected node to a source node to handle the migration of the task.
(2) *Task Migration Phase*. During this phase, the source node dispatches the cloned task image and data to a target node, to be restored and resumed.
(3) *Resuming and Completion Phase*. As soon as the task information starts to arrive, the target node starts to restore the task. When all the information has been received, the execution of the task is resumed on the target node.

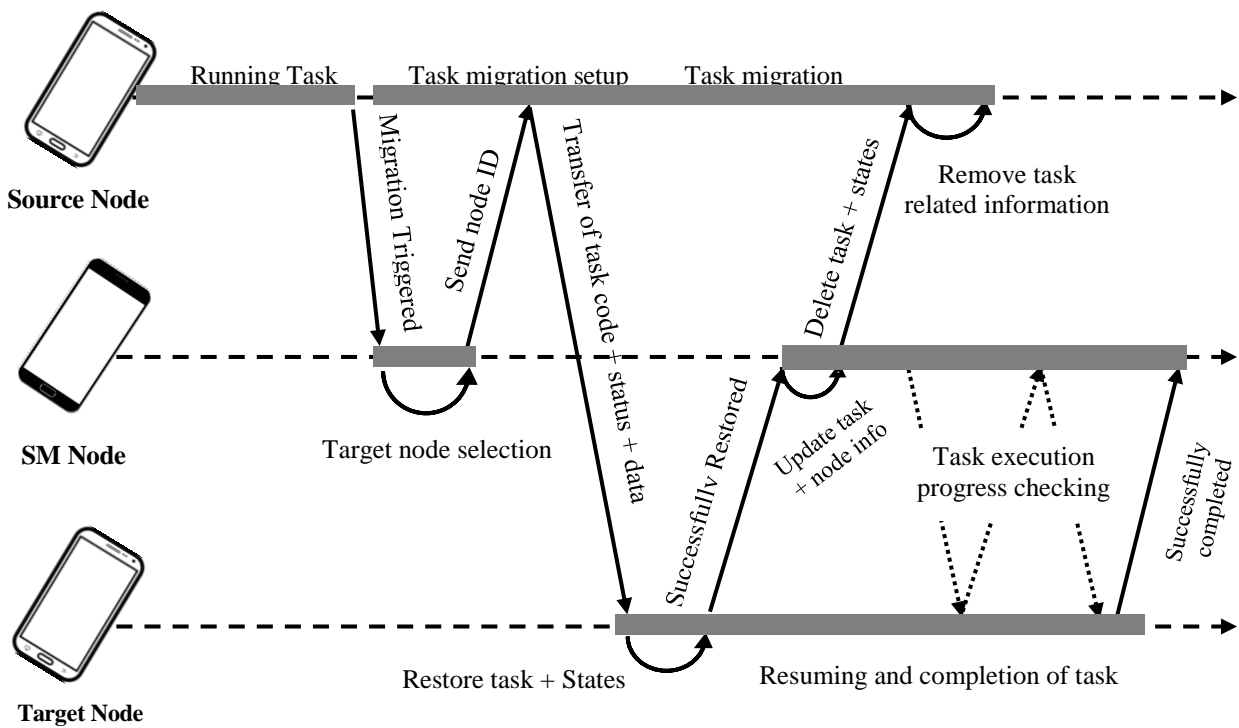

**Figure 7. Phases of the migration process.**

The step-by-step process of migrating a task (Fig. 8) is as follows:

(1) SPN1 monitors the node and task progress and triggers the migration process by sending a request to the resource allocation service deployed on the SMN. The request includes the ID of the existing SPN and task information.
(2) The resource allocation service deployed on the SMN selects another SPN2 to execute the task.
(3) The ID of the selected node is sent to the dispatcher on SPN1.
(4) The task code, related data, and current status are sent to SPN2.
(5) The task resumes execution on SPN2.
(6) Upon successful completion of the task, the result is sent to the SCN and the status is updated in the task queue deployed on the SMN.

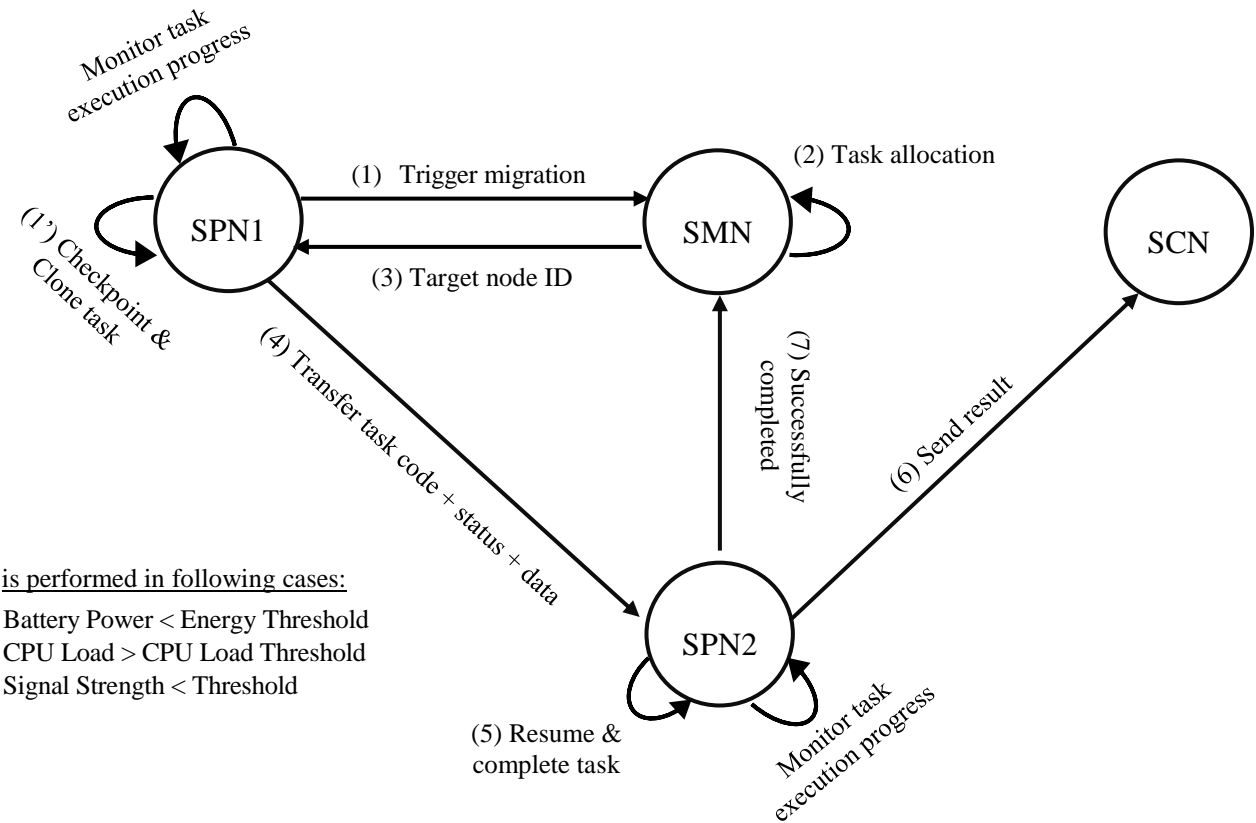

**Figure 8.** Step-by-step process for migration of a task.

## C. WORKING of RMS

In this section, we briefly describe the workings of the proposed RMS. Detailed mechanisms employed by each component are discussed in the related sections.

The user node or SCN submits task information to the task queue via the task queue manager. The node- and network-level information collected by the discovery and monitoring manager is stored in the resource pool. The resource allocation service accesses task information from the task queue and node information from the resource pool and selects an SPN that fulfills the task requirements. The IDs of the task and selected SPN are sent to the dispatcher deployed on the SCN, which transfers the task code and related data to the selected SPN. The dispatcher deployed on the SPN facilitates the execution of the received task and, upon successful completion, sends the result to the SCN and updates the status at the resource pool and the task queue.

## IV. PRELIMINARY RESULTS

In order to demonstrate the feasibility of the proposed system, Contiki [33] was used to implement the resource allocation scheme and the Cooja [34, 35] simulator was used to implement network layer functions. Altogether, 20 nodes were randomly deployed, and numerous tasks of varying sizes were used to reflect data-intensive applications, such as distributed face recognition. Nodes were divided into three categories: SMNs, SPNs, and SCNs. The SMN was responsible for allocating tasks submitted by SCNs. Based on the resource allocation policy, SPNs were selected to execute the tasks. Cooja's mobility plugin was used to generate group mobility patterns. The simulation parameters are given in Table 2, and the network scenario is shown in Fig. 9.

The performance of the proposed scheme is compared with that of a heterogeneity-aware task allocation (HTA) scheme [51], which allocates tasks on the basis of node processing power and available energy.

**Table II.** Simulation parameters.

| Platform | Cooja Simulator Contiki 3.0 |
|---|---|
| Simulation time | 3,600 s |
| Number of nodes | 20 |
| Mote types | 3 |
| Simulation area | 1,200 m × 1,200 m |
| Packet size | 512 bytes |
| MAC | IEEE 802.11 |
| Transport protocol | User Datagram Protocol (UDP) |
| Radio medium | UDGM [34] |
| Transmission range | 180 m |
| INT range | 20 m |

In HTA, nodes use the maximum transmission power for communication. The scheme also does not consider network-level information, such as link quality and lifetime.

The performance is evaluated in terms of accumulative task completion time (ATCT):

$$ATCT = \sum_{i=1}^{n} E_{CT}(T_i). \qquad (15)$$

The completion time for task $E_{CT}(T_i)$ consists of the CPU execution time and the data transfer time. The task completion time estimation model is described in Section III.

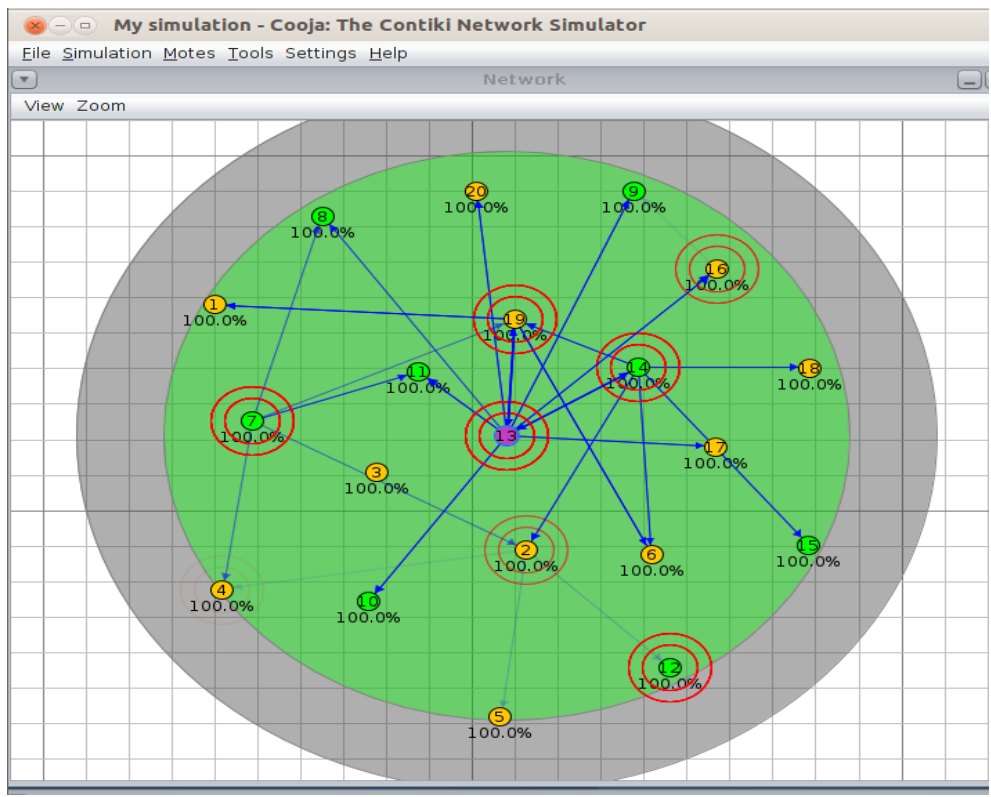

Figure 9. **Network scenario.**

Scenario 1. The proposed scheme makes allocation decisions on the basis of the estimated task completion time and energy consumption. The task completion time comprises the task execution time and the data transfer time. The task execution time includes the CPU processing time, queue time, and system overhead. The simulation results in Fig. 10 show that the performance of the proposed scheme is better by 10–15%. With 10 tasks, there is a minor improvement, but as the number of tasks increases, the performance of the scheme improves. The HTA scheme performs poorly because it does not consider data transfer costs. It selects nodes with the highest processing power and available energy. In several cases, nodes with the highest processing power and available energy were connected via low-quality communication links, which increased the data transfer time and thus the task completion time.

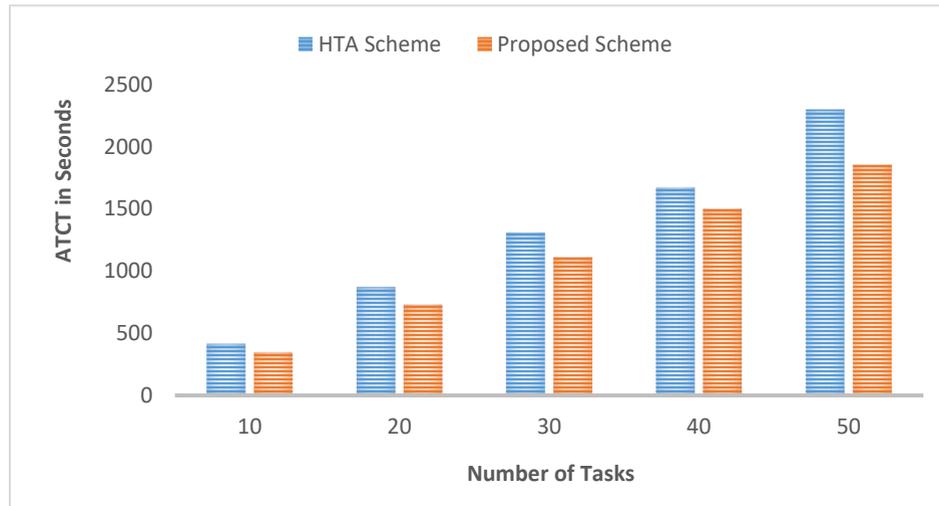

**Figure 10.** ATCT for Scenario 1.

Scenario 2. The proposed scheme makes allocation decisions on the basis of the estimated task completion time, energy consumption, and link lifetime. The results for Scenario 2 in Fig. 11 show that, for the proposed scheme, the performance improves by 18–34%. This is because the proposed scheme selects SPNs with a route lifetime greater than or equal to the estimated data transfer time of tasks. This reduces route rediscovery and reselection costs in case of link failure or global node mobility. The transmission power control mechanism increases the number of parallel transmissions and thus the network capacity. This reduces the overall data transfer cost and thus the ATCT. HTA does not consider link lifetime and transmission power, so it select nodes with an unstable connectivity and reduced network capacity.

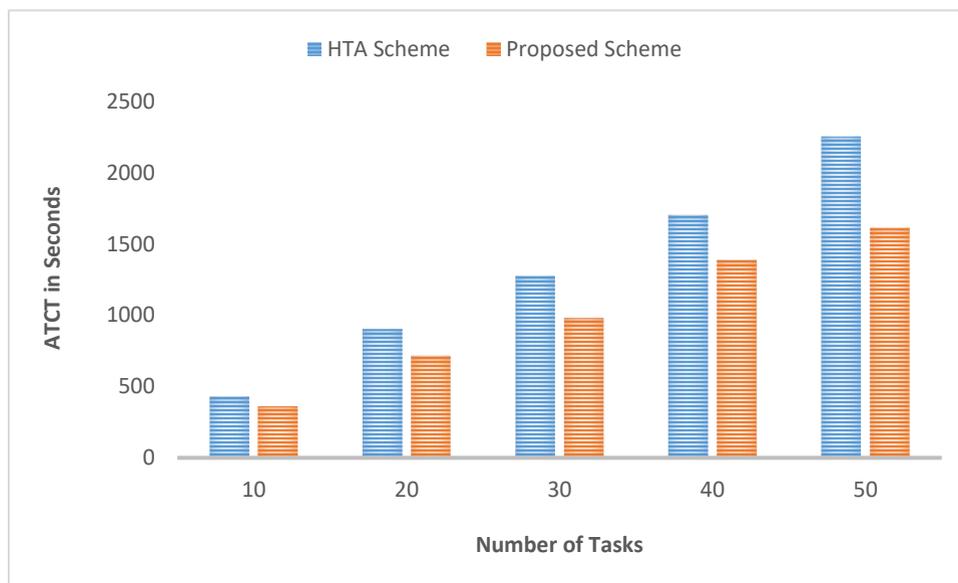

**Figure 11.** ATCT for Scenario 2.

*Transmission Energy Consumption*
The transmission energy consumption of the proposed scheme was compared with that of a minimum hop-based task allocation (MinHop) scheme [9] as well as HTA [51]. The MinHop scheme assigns a task to a node on the basis of the number of hops. It uses the maximum transmission power. In the proposed scheme, nodes use three transmission power levels for communication. Two scenarios were defined. In both scenarios, 20 nodes were deployed. However, the distance between the nodes in Scenario 3 was uniform, whereas in Scenario 4, nodes were deployed in groups of numerous sizes.

The transmission energy consumption for Scenarios 3 and 4 is given in Figs. 12 and 13, respectively. The performance of the proposed scheme in Scenario 3 is poor because most of the nodes were accessible only at the maximum transmission power. So, like MinHop and HTA, in the proposed scheme, the nodes used the maximum transmission power for communication. In addition, because of the multiple transmission power levels, the amount of control traffic in the proposed scheme was three times more than in the other schemes.

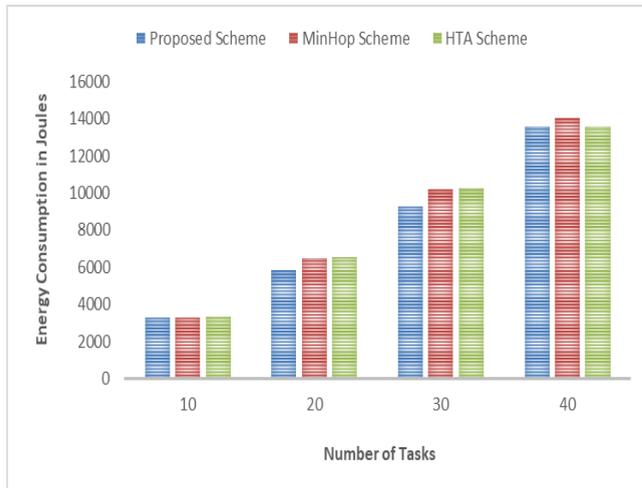

Figure 12. Transmission energy consumption for Scenario 3.

In Scenario 4, the proposed scheme outperforms the other two schemes. This is because nodes were deployed in groups. Several nodes were accessible at lower transmission power levels. The proposed scheme allocated tasks to nodes accessible at the minimum transmission power, which significantly reduced the transmission energy.

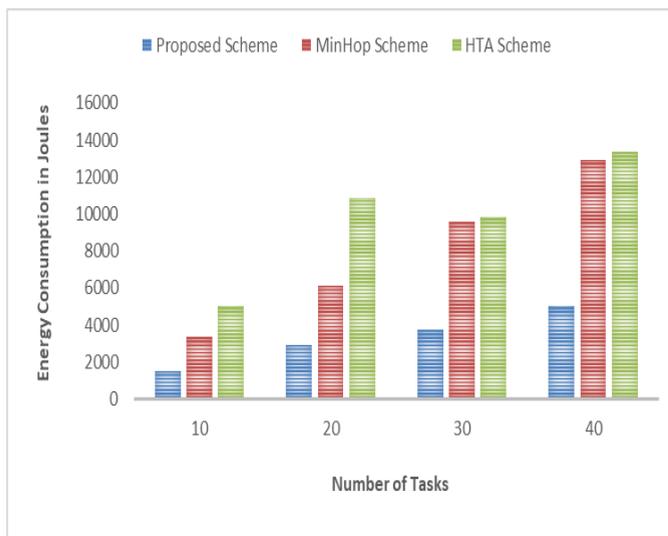

Figure 13. Transmission energy consumption for Scenario 4.

The transmission energy consumption of the proposed scheme using three transmission power levels compared to the proposed scheme always using the maximum transmission power is given in Fig. 14. The results indicate that the transmission power control mechanism significantly reduces the transmission energy.

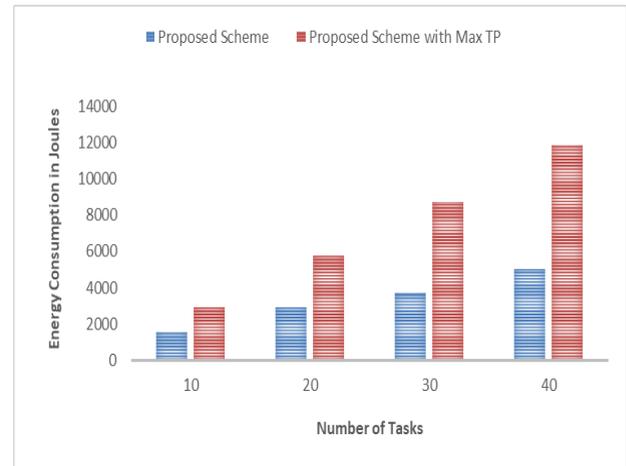

Figure 14. Transmission energy consumption for Scenario 4.

## V. CONCLUSION

RMSs are responsible for the discovery, monitoring, migration, and allocation of network and system resources. They are an integral part of mobile ad hoc clouds and play a key role in the application and system performance. Research into resource management for mobile ad hoc clouds is still in a preliminary phase, and very few schemes based on a decentralized architecture have been proposed in order to address issues such as node mobility, energy management, and task failure. Most of these schemes do not consider the network environment, task queue size, or CPU overhead. Very few schemes consider the network environment, but they do not consider link quality, link lifetime, or the migration and reallocation of tasks.

In this paper, we proposed an energy-efficient RMS to support the execution of mobile CPS applications on a mobile ad hoc cloud. The proposed system consisted of two layers: a network layer and a middleware layer. The network layer provides ad hoc network and communication services to the middleware layer and shares the collected information in order to enable efficient and robust resource management decisions. The middleware layer is responsible for the discovery, monitoring, migration, and allocation of resources.

The new system is different from the systems proposed in the literature because it focuses on mobile CPS applications and energy-efficient communication between tasks. In addition, it aims to address task failure by adopting a failure avoidance mechanism, and it uses a cross-layer design to support effective resource management decisions. Our research on transmission power control mechanisms is used to reduce the transmission energy consumption and increase concurrent

VOLUME XX, 2017                                                                                                                                                                                                          9

transmissions in the network. The new system considers the processing power, queue size, system overhead, link quality, and link lifetime to reduce execution times, data transfer costs, and energy consumption.

## ACKNOWLEDGMENTS
I would like to acknowledge contribution of Mr. Mpyana Mwamba Merlec of Korea University.

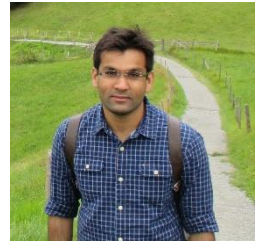

**Sayed Chhattan Shah** is an Assistant Professor of computer science at the Department of Information Communication Engineering at Hankuk University of Foreign Studies, Seoul, South Korea. He is also the Director of the Mobile Grid and Cloud Computing Laboratory. His research interests lie in the fields of parallel and distributed computing systems, mobile computational clouds, and ad hoc networks.

He received his Ph.D. degree in computer science from Korea University in 2012 and his M.S. degree in the same field from the National University of Computer and Emerging Sciences in 2008. Prior to joining HUFS, he was a Senior Researcher at the Electronics and Telecommunications Research Institute, Republic of Korea, and an Engineer at the National Engineering and Scientific Commission, Pakistan. He also held faculty positions at Seoul National University of Science and Technology, Korea University, Dongguk University, Hamdard University, and Isra University.

Shah is an Associate Editor of Information Processing Systems and Intelligent Automation and Soft Computing. He has served as the Conference Chair and on program committees of various international conferences. He is a Senior Member of IEEE and a Member of IEEE Communications Society, International Telecommunication Union, and the European Research Council.